\documentstyle[12pt]{article}

\newcommand{\be}{\begin{equation}}
\newcommand{\ee}{\end{equation}}
\newcommand{\bea}{\begin{eqnarray}}
\newcommand{\eea}{\end{eqnarray}}
\newcommand{\bx}{\mbox{\boldmath $x$}}
\newcommand{\bS}{\mbox{\boldmath $S$}}
\newcommand{\tbS}{\mbox{\hskip 3pt $\tilde{\mbox{\hskip -3pt $\bS$}}$}}
\newcommand{\bI}{\mbox{\boldmath $I$}}
\newcommand{\bpartial}{\mbox{\boldmath $\partial$}}
\newcommand{\bD}{\mbox{\boldmath $D$}}

\title{
\hfill{\normalsize ULB/229/CQ/98/7; IOP-BBSR/98-26}\\
\vspace{1cm}
On some one-parameter families of three-body problems in one
dimension: Exchange operator formalism in polar coordinates and scattering
properties}
\author{Avinash Khare $^{a,}$\thanks{E-mail: khare@iopb.stpbh.soft.net}\ ,
C. Quesne
$^{b,}$\thanks{Directeur de recherches FNRS; E-mail: cquesne@ulb.ac.be} \\
{\small
\sl $^a$ Institute of Physics, Sachivalaya Marg, Bhubaneswar 751005, India}\\
{\small \sl $^b$ Physique Nucl\'eaire Th\'eorique et Physique Math\'ematique,
Universit\'e Libre de Bruxelles,} \\ {\small \sl Campus de la Plaine CP229,
Boulevard~du Triomphe, B-1050 Brussels, Belgium}}
\date{}

\begin{document}
\maketitle
\begin{abstract}  We apply the exchange operator formalism in polar
coordinates to
a one-parameter family of three-body problems in one dimension and prove the
integrability of the model both with and without the oscillator potential.
We also
present exact scattering solution of a new family of three-body problems in one
dimension.
\end{abstract}

\vspace{0.5cm}

\hspace*{0.3cm} PACS: 03.65.-w, 03.65.Ge, 03.65.Fd

\hspace*{0.3cm} Keywords: Three-body problems, exchange operators, scattering
\newpage
%
%
In recent years, the Calogero-Sutherland type~\cite{1,2} $N$-body problems in
one dimension have attracted a lot of attention~\cite{3}. Some time ago,
Brink et
al.~\cite{4}  and Polychronakos~\cite{5} independently introduced an exchange
operator formalism, leading to covariant derivatives, known in the mathematical
literature as Dunkl operators~\cite{6}, and an $S_N$-extended Heisenberg
algebra~\cite{7,8}.  In terms of this formalism, the $N$-body quantum Calogero
model has been shown to be equivalent to a  set of free modified oscillators and
hence integrable.\par
%
%
In the last few years, there has been renewed interest in three-body
Calogero-Marchioro-Wolfes (CMW) problem~\cite{9,10}, and some other related
three-body problems~\cite{11,12}, all of which include a three-body
potential. It is
then natural to enquire if the exchange operator formalism~\cite{4,5} can
also be
extended to these problems and further if using it, one can show the
integrability of
these models. Recently, one of us~(CQ) took the first step in that
direction when she
showed the  integrability of the quantum CMW problem by using the exchange
operator formalism~\cite{13}.\par
%
%
The purpose of this letter is to extend the exchange operator formalism to the
class of three-body problems, with and without the oscillator potential, discovered
recently by Sukhatme and one of the author (AK)~\cite{14}, and hence to
prove the
integrability of the quantum model. It is worth pointing out that in this
model there
is an interesting relationship between the incoming and outgoing momenta of the
three particles. In particular, we show that introducing an exchange operator
formalism in polar coordinates is very useful in analyzing this model (both with
and without the oscillator potential) and proving its integrability. Finally, we
discuss the scattering solution for a new one-parameter family of three-body
problems and show that even for these problems there is a very simple
relationship
between the incoming and outgoing momenta of the particles.\par
%
%
The exchange operator formalism in polar coordinates has not been discussed so
far in the literature, hence it may be worthwhile to first discuss the CMW
three-body problem (a known integrable model~\cite{13}) using this formalism. We
shall see that the generalization of the formalism to the class of three-body
problems to be  discussed below~\cite{14} is then straightforward.\par
%
%
The three-particle Hamiltonian for the CMW problem is given by~\cite{9,10}
\be \label{eq:CMW-H}
  H = \sum^3_{j=1} \left(-\partial_j^2+\omega^2 x^2_j\right)
  +g\sum^3_{\scriptstyle i,j=1 \atop \scriptstyle {i\not =j}}{1\over
  (x_i-x_j)^2} + 3f \sum^3_{\scriptstyle i,j,k=1 \atop \scriptstyle
  {i\ne j\ne k\ne i}} {1\over (x_i+x_j-2x_k)^2},
\ee
where $x_i$ ($i=1$, 2, 3) denotes the particle coordinates, $\partial_i\equiv
\partial/\partial x_i$, and the inequalities $g$, $f > - 1/4$ are assumed
to prevent
collapse. Let $x_{ij}\equiv x_i-x_j$ and $y_{ij}\equiv x_i+x_j-2x_k$ ($i\not =
j\not = k\not = i$), where in the latter we suppress index $k$ since it is
entirely
determined  by $i$ and $j$.\par
%
%
Let us introduce the Jacobi coordinates
\be \label{eq:Jacobi}
  R = {x_1+x_2+x_3\over 3}, \qquad x \equiv {x_{12}\over\sqrt 2} = r \sin\phi,
  \qquad y \equiv {y_{12}\over \sqrt 6} = r \cos\phi.
\ee
It is then easily verified that $r^2={1\over 3} \sum^3_{i<j} (x_i-x_j)^2$, and
\be \label{eq:x-y}
  x_{ij} = \sqrt 2\, r \sin \left(\phi +{2\pi k\over 3}\right), \qquad
  y_{ij} = \sqrt 6\, r \cos \left(\phi +{2\pi k\over 3}\right),
\ee
where $(ijk) = (123)$. One can show that the differential operators
$\partial_R$,
$\partial_r$ and $\partial_{\phi}$ are given in terms of $\partial_i$
($i=1$,  2, 3)
by
\be \label{eq:partial}
  \partial_R = \sum_j \partial_j, \qquad \partial_r = - \sqrt{{2\over 3}} \sum_j
  \cos \left(\phi + {2\pi\over 3}j\right) \partial_j, \qquad
  \partial_{\phi}=\sqrt{{2\over 3}}\, r\sum_j \sin \left(\phi + {2\pi\over
  3}j\right)\partial_j.
\ee
As a result, in polar coordinates the CMW Hamiltonian (\ref{eq:CMW-H}) takes the
form $H = H_R+H_r$, where
\be \label{eq:H_R}
  H_R = - {1\over 3} \partial^2_R+3\omega^2 R^2,
\ee
\be \label{eq:H_r}
  H_r = - \partial^2_r - {1\over r} \partial_r - {1\over r^2}\partial^2_{\phi}+
  \omega^2r^2+{9\over r^2} \left({g\over \sin^2 3\phi}+{f\over \cos^2
3\phi}\right).
\ee
\par
%
%
In Ref.~\cite{13}, the CMW problem was analyzed in terms of some exchange
operators belonging to a $D_6$-group. The latter is generated by the particle
permutation operators~$K_{ij}$, and the inversion operator~$I_r$ in relative
coordinate space. Let us now consider the effect of $K_{ij}$ and $I_r$ on
the polar
coordinates $R$, $r$, $\phi$. Using the fact that
\be
  K_{ij} x_{j} = x_{i}K_{ij}, \qquad K_{ij} x_{k} = x_{k}K_{ij}, \qquad I_r
x_i = (2
  R-x_i) I_r,
\ee
it is easy to show that
\bea
  K_{ij} R & = & R K_{ij}, \qquad K_{ij} r = r K_{ij}, \qquad K_{ij}\phi =
\left(-\phi
  +{2\pi\over 3}k\right) K_{ij}, \nonumber \\
  I_r R & = & RI_r, \qquad I_r r = rI_r, \qquad I_r\phi = (\phi+\pi)I_r.
\eea
Hence it follows that
\be
  L_{ij} R = R L_{ij}, \qquad L_{ij} r = r L_{ij}, \qquad L_{ij} \phi =
\left[-\phi
  + (2k+3){\pi\over 3}\right] L_{ij},
\ee
where $L_{ij} \equiv K_{ij}I_r$. We thus see that $K_{ij}$, $I_r$, and hence all
the operators of the $D_6$-group act only on $\phi$ (and not on $r$ and $R$).
Furthermore, the operations may be written in terms of the two operators
$\cal R$ and $\cal I$, defined by
\be \label{eq:rot-inv}
  {\cal R} = \exp \left({\pi\over 3}\partial_{\phi}\right), \qquad
  {\cal I} = \exp (\mbox{\rm i}\pi \phi \partial_{\phi}),
\ee
where $\cal R$ and $\cal I$ denote the rotation operator by angle $\pi/3$
and the
inversion operator, respectively, i.e.,
\be
  {\cal R} \psi (\phi) = \psi (\phi +\pi/3) {\cal R}, \qquad
  {\cal I} \psi (\phi) = \psi (-\phi) {\cal I}.
\ee
In particular, in terms of $\cal R$ and $\cal I$ the 12 generators of the
$D_6$-group are given by
\bea \label{eq:D_6}
  & &I, \qquad K_{ij} = {\cal I} {\cal R}^{2k}, \qquad K_{123} = {\cal
R}^2, \qquad
       K_{132} = {\cal R}^4, \nonumber \\
  & &I_r= {\cal R}^3, \qquad L_{ij} = {\cal I}{\cal R}^{2k+3}, \qquad
L_{123} = {\cal
       R}^5, \qquad L_{132} = {\cal R},
\eea
where $K_{123} \equiv K_{12} K_{23}$, $K_{132} \equiv K_{23} K_{12}$, $L_{123}
\equiv K_{123} I_r$, and $L_{132} \equiv K_{132} I_r$.\par
%
%
{}Following Eq.~(\ref{eq:partial}), it is natural to define the covariant
derivatives
$D_R$, $D_r$, and $D_{\phi}$ in polar coordinates by
\be \label{eq:D-polar}
  D_R = \sum_j D_j, \qquad D_r = -\sqrt{{2\over 3}} \sum_j
  \cos \left(\phi + {2\pi\over 3}j\right) D_j, \qquad
  D_{\phi} =\sqrt{{2\over 3}}\, r \sum_j \sin\left(\phi +{2\pi\over
3}j\right)D_j,
\ee
where $D_i$ ($i=1$, 2, 3) are those in cartesian coordinates.\par
%
%
On using the fact that in CMW model the generalized derivatives are defined
by~\cite{13}
\be \label{eq:D}
  D_i = \partial_i - \kappa \left({1\over x_{ij}} K_{ij}-{1\over x_{ki}}
K_{ki}\right)
  - \lambda \left({1\over y_{ij}} L_{ij}+{1\over y_{ki}} L_{ki}-{2\over
y_{jk}} L_{jk}
  \right),
\ee
where $g = \kappa(\kappa-1)$, and $f=\lambda(\lambda-1)$, $D_R$, $D_r$, and
$D_{\phi}$ can be shown to be given by
\bea \label{eq:D-polar-bis}
  D_R & = & \partial_R, \qquad D_r = \partial_r - {\kappa\over r}
\left(\sum_j {\cal
           R}^{2j}\right) {\cal I} - {\lambda\over r} \left(\sum_j {\cal
R}^{2j+1}\right)
           {\cal I}, \nonumber \\
  D_{\phi} & = & \partial_{\phi} - \kappa \sum_j \cot \left[\phi+(3-j)
{2\pi\over
           3}\right] {\cal R}^{2j} {\cal I} \nonumber \\
  & & \mbox{} +\lambda \sum_j \tan \left[\phi+(4-j){2\pi\over 3}\right] {\cal
           R}^{2j+1} {\cal I}.
\eea
It is easy to show that the covariant derivatives for the pure three-body case
($\kappa=0$)  can be obtained from those for the pure two-body one ($\lambda=0$)
by making the  rotation $\phi \rightarrow \phi' = \phi +\pi/6$. In the
proof, one
uses the fact that the operators ${\cal R}'$ and ${\cal I}'$, defined in
terms of
$\phi'$ in the same way as $\cal R$ and $\cal I$ in terms of $\phi$ (see
Eq.~(\ref{eq:rot-inv})), can be expressed in terms of the latter as ${\cal
R}' = {\cal
R}$, and ${\cal I}' = \cal RI$, respectively.\par
%
%
Let us now consider the exchange operator formalism for the one-parameter
family of three-body problems~\cite{14} characterized by
\be \label{eq:genCMW-H}
  H =\sum^3_{j=1} (-\partial^2_j+\omega^2 x^2_j)
  +g \sum^3_{\scriptstyle i,j=1 \atop \scriptstyle {i \ne j}} {1\over
  x'^2_{ij}}+ 3f \sum^3_{\scriptstyle i,j=1 \atop \scriptstyle {i \ne j}}
  {1\over y'^2_{ij}},
\ee
where
\be \label{eq:gen-x-y}
  x'_{ij}\equiv x_{ij} \cos\delta +{y_{ij}\over \sqrt 3} \sin\delta, \qquad
  y'_{ij} = -\sqrt 3\, x_{ij} \sin\delta +y_{ij} \cos\delta,
\ee
and $0 \le \delta \le \pi/6$. On following Eqs.~(\ref{eq:Jacobi}) to (\ref{eq:H_R}),
the Hamiltonian (\ref{eq:genCMW-H}) takes the form $H= H_R +H'_r$, where
\be
  H'_r = -\partial^2_r-{1\over r} \partial_r - {1\over r^2} \partial^2_{\phi}
  +\omega^2 r^2+{9\over r^2} \bigg [{g\over \sin^2(3\phi+3\delta)}+{f\over
  \cos^2(3\phi+3\delta)} \bigg ],
\ee
while $H_R$ is again as given by Eq. (\ref{eq:H_R}).\par
%
%
Now notice that H for this problem can be obtained from the CMW Hamiltonian
(Eqs. (\ref{eq:H_R}),~(\ref{eq:H_r})) by making the change of variables
\be
  R \rightarrow R' = R, \qquad r \rightarrow r' = r, \qquad \phi\rightarrow
\phi'
  = \phi +\delta.
\ee
Hence the exchange operator formalism developed above for the CMW case
remains valid for even this case, provided we replace all coordinates and
operators
by the corresponding primed ones. For example, by using
Eq.~(\ref{eq:D-polar-bis}),
the generalized derivatives in the primed polar coordinates are given by
\bea \label{eq:genD-polar}
  D'_R & = & D_R = \partial_R, \qquad D'_r = \partial_r  -{\kappa\over r} \left(
         \sum_j {\cal R}'^{2j}\right) {\cal I'} - {\lambda\over r}
\left(\sum_j {\cal
         R}'^{2j+1}\right) {\cal I'}, \nonumber \\
  D'_{\phi} & = & \partial_{\phi} - \kappa \sum_j \cot \left[
         \phi+\delta+(3-j){2\pi\over 3}\right] {\cal R'}^{2j} {\cal I'}
\nonumber \\
  & & \mbox{} +\lambda \sum_j \tan \left[\phi+\delta+(4-j){2\pi\over 3}\right]
         {\cal R'}^{2j+1} {\cal I'},
\eea
where ${\cal R'}$, ${\cal I'}$ are again defined in terms of $\phi'$ as
${\cal R}$,
${\cal I}$ in terms of $\phi$. From Eq.~(\ref{eq:D-polar}), it follows that
$D'_R$,
$D'_r$, $D'_{\phi}$ can be expressed  in terms of generalized derivatives
$D'_i$ in some primed coordinates $x'_i$ by
\bea
  D'_R & = & \sum_j D'_j, \qquad D'_r = - \sqrt{{2\over 3}} \sum_j \cos
         \left(\phi+\delta+ {2\pi\over 3}j\right) D'_j, \nonumber \\
  D'_{\phi} & = & \sqrt{{2\over 3}}\, r \sum_j \sin
\left(\phi+\delta+{2\pi\over 3}j
         \right) D'_j.
\eea
Note that $x'_i$ is defined in terms of $R'(=R)$, $r'(=r)$, $\phi'$ in the
same way
as
$x_i$ in terms of $R$, $r$, $\phi$.\par
%
%
{}Further, following Eq.~(\ref{eq:D}), $D'_i$ can also be written as
\be
  D'_i = \partial'_i - \kappa \left({1\over x'_{ij}} K'_{ij}-{1\over x'_{ki}}
  K'_{ki}\right) - \lambda \left({1\over y'_{ij}} L'_{ij}+{1\over y'_{ki}}
  L'_{ki}-{2\over y'_{jk}} L'_{jk}\right),
\ee
in terms of the primed coordinates $x'_i$, and some primed operators $K'_{ij}$,
$L'_{ij}$. The latter have the same action on $x'_k$ as $K_{ij}$, $L_{ij}$
on $x_k$,
respectively, and belong to a tranformed $D_6$-group. The operators $D'_i$ are
covariant under this transformed $D_6$-group, and commute with one another
(hence they may be called Dunkl operators~\cite{6}).\par
%
%
We now need to determine the relation between $x'_i$ and $x_i$ to study the
action of the transformed $D_6$-group on $x_i$. To this end we start from the
relation
\be
  x'_k = - {1\over 3} y'_{ij}+R = - {1\over\sqrt 3} \left( - x_{ij} \sin\delta
  + {y_{ij}\over\sqrt 3} \cos\delta \right) +R.
\ee
On using Eqs.~(\ref{eq:Jacobi}) and~(\ref{eq:x-y}), we then have
\be
  x'_k = s_2 x_i + s_1 x_j+s_3 x_k, \qquad (ijk) = (123),
\ee
where
\be
  s_k \equiv s_k (\delta) = {1\over 3} \left[1+2 \cos\left(\delta+{2\pi\over 3}k
  \right)\right].
\ee
In a compact matrix form, we can write $\bx' = \bx \bS$, where $\bx = (x_1 x_2
x_3)$, $\bx' = (x'_1 x'_2 x'_3)$, and the matrix $\bS$ is given by
\be
  \bS = \bS(\delta) = \pmatrix{ s_3 & s_1 & s_2\cr s_2 & s_3 & s_1\cr
  s_1 & s_2 & s_3\cr}.
\ee
It is easily checked that $\bS$ is orthogonal, i.e., $\bS \tbS = \tbS \bS =
\bI$, hence we also have
\be \label{eq:S-transf}
  \bx = \bx' \tbS, \qquad \bpartial'  = \bpartial \bS, \qquad \bpartial =
  \bpartial' \tbS,
\ee
where $\bpartial = (\partial_1 \partial_2 \partial_3)$, $\bpartial' = (\partial'_1
\partial'_2 \partial'_3)$. Thus, by definition $\bS$ is the matrix
representing the
operator ${\cal S} = \exp (\delta \partial_{\phi})$ of rotation through an
angle~$\delta$.\par
%
%
Under the transformation $\phi \rightarrow \phi' = \phi+\delta$, the functions
transform as
\be
  \psi (\phi) = \psi (\phi'-\delta) = \psi'(\phi'),
\ee
hence it is obvious that ${\cal R'} = {\cal R}$, while ${\cal I'} \equiv \exp
(\mbox{\rm i}\pi\phi'\partial'_{\phi})$ is such that
\be
  {\cal I'} \psi(\phi) = {\cal I'} \psi'(\phi') = \psi' (-\phi') = \psi
(-\phi'-\delta)
  = \psi(-\phi-2\delta)= {\cal S}^2 {\cal I} \psi (\phi).
\ee
In other words, ${\cal I'} = {\cal S}^2{\cal I}$, where ${\cal S}^2$ is the
operator
of rotation through an angle~$2\delta$. Using these results, the operators
$D'_R$, $D'_r$, and $D'_{\phi}$ of Eq.~(\ref{eq:genD-polar}) may be expressed in
terms of unprimed variables and operators. On the other hand, from
Eq.~(\ref{eq:D_6}) and its primed counterpart, it follows that the
operators of the
transformed $D_6$-group may be written as
\bea \label{eq:transf-D_6}
  & &I, \qquad K'_{ij} = {\cal S}^2 K_{ij}, \qquad K'_{123} = K_{123}, \qquad
          K'_{132} = K_{132}, \nonumber \\
  & &I'_r = I_r, \qquad L'_{ij} = {\cal S}^2 L_{ij}, \qquad L'_{123} =
L_{123}, \qquad
          L'_{132} = L_{132}.
\eea
{}From these relations, their action on $x_k$ can be easily determined, but
will not
be given in detail here.\par
%
%
{}Finally, we may introduce a Hamiltonian with exchange terms
\be \label{eq:H_exch}
  H_{exch} = \sum^3_{j=1} \left(-\partial_j^2 + \omega^2 x_j^2\right) +
  \sum^3_{\scriptstyle i,j=1 \atop \scriptstyle i \ne j} {1\over x'^2_{ij}}
  \kappa(\kappa-K'_{ij}) + 3 \sum^3_{\scriptstyle i,j=1 \atop \scriptstyle
i\ne j}
  {1\over y'^2_{ij}}\lambda(\lambda-L'_{ij}).
\ee
In case $\omega=0$, since $\cal S$ is an orthogonal matrix, it is easily shown
that
\be
  H_{exch} = - \sum^3_{j=1} D'^2_j.
\ee
It is worth noting that $H_{exch}$ may also be written as $H_{exch} = -
\sum_{j=1}^3 D_j^2$ in terms of Dunkl operators in unprimed coordinates, defined
as in Eq.~(\ref{eq:S-transf}) by $\bD = \bD' \tbS$. Further, the operators
$I_n = \sum_{i=1}^3 \Pi'^{2n}_i$ ($n=1$, 2, 3), where $\Pi'_i = - \mbox{\rm
i} D'_i$
are generalized momenta, commute with one another, and are left invariant under
the transformed $D_6$-group. Hence, their projection in the subspaces of Hilbert
space characterized by $(K'_{ij},L'_{ij}) = (1,1), (1,-1), (-1,1)$ or
$(-1,-1)$ also commute. In these subspaces, $I_1 = H_{exch}$ is nothing but the
one-parameter family of Hamiltonians as given by Eq.~(\ref{eq:genCMW-H}) with
$\omega = 0$, corresponding to the parameter values $(\kappa,\lambda),
(\kappa,\lambda+1), (\kappa+1,\lambda)$ or $(\kappa+1,\lambda+1)$,
respectively,
while $I_2$ and
$I_3$ become the  integrals of motion of such Hamiltonians.\par
%
%
In case $\omega \not =0$, i.e., when the oscillator potential is present, the
Hamiltonian (\ref{eq:H_exch}) can be written as
\be \label{eq:H_exch-bis}
  H_{exch} = \sum^3_{i=1} ( -D'^2_i+ \omega^2 x'^2_i)  = \omega \sum^3_{i=1}
\{a'^{+}_i, a'_i\},
\ee
where
\be \label{eq:a}
  a'_i = {1\over\sqrt{2\omega}} (\omega x'_i+ D'_i), \qquad
  a'^{+}_i = {1\over\sqrt{2\omega}} (\omega x'_i-D'_i).
\ee
One can now show that $a'_i$, $a'^+_i$ ($i=1$, 2, 3), and the operators of the
transformed $D_6$-group, as given by Eq.~(\ref{eq:transf-D_6}), generate  a
$D_6$-extended oscillator algebra, whose commutation relations are entirely
similar to Eq.~(4.2) of Ref.~\cite{13}. It may also be noted that since
$\bS$ is an orthogonal matrix, $H_{exch}$ as given by Eq.~(\ref{eq:H_exch})
can also
be written in terms of $D_i$, $x_i$, or $a_i$, $a^+_i$ defined exactly  as in
Eqs.~(\ref{eq:H_exch-bis}) and~(\ref{eq:a}). Thus we see that
Hamiltonian~(\ref{eq:genCMW-H}) corresponding to the parameter values
$(\kappa,\lambda)$, $(\kappa,\lambda+1)$, $(\kappa+1,\lambda)$ or
$(\kappa+1,\lambda+1)$ can be regarded as a free modified boson Hamiltonian. The
corresponding conserved quantities are
\be
  I_n = \sum^3_{i=1} h'^n_i, \qquad \ h'_i = {1\over 2} \{a'^+_i,a'_i\},
\qquad n =
  1,2,3,
\ee
and following Ref.~\cite{13}, it is easily shown that $I_1$, $I_2$, $I_3$ are
mutually commuting operators. Note that $I_n$, $n=1$, 2, 3, are invariant
under the
transformed $D_6$-group, and hence their projections in the subspaces
characterized by $K'_{ij}$ and $L'_{ij}$ equal to $+1$  or $-1$ still
commute with
one another.\par
%
%
Before ending this note, we would like to present a new one-parameter family of
three-body problems in one dimension, and show that there is an interesting,
simple relation between the incoming and outgoing momenta of the three
particles.
Since the philosophy is similar to that of Ref.~\cite{14}, we avoid giving
all the
details here, but merely point out the steps that are different in the two
cases. As
in Ref.~\cite{14}, we work in the centre-of-mass frame and consider the
following
relative Hamiltonian in polar coordinates $(\hbar = 2m = 1)$
\be \label{eq:newH}
  H = - \left({\partial^2\over\partial r^2}
  + {1\over r}{\partial\over\partial r}\right)
  + {B^2\over  r^2},
\ee
where
\be \label{eq:B}
  B^2 = - {\partial^2\over\partial\phi^2}
  + {9d^2g\over \sin^2 (3d\phi)},
\ee
with $0\leq \phi \leq \pi/3d$, and $d = 1$, 2, 3. Note that for $d = 1$, we
obtain
the famous Calogero problem, while for other $d$ values, we get some new
three-body problems. For $d=2$, for instance, the potential in
Eq.~(\ref{eq:newH}) is
$V = 12g r^2 \sum_{i<j}^3 x_{ij}^{-2} y_{ij}^{-2}$. The angular Schr\"odinger
equation is easily solved by following Calogero~\cite{1}, yielding
\be
  B_l = 3d ( l + \kappa ), \qquad g = \kappa(\kappa-1).
\ee
\par
%
%
If $d$ is even, then on running through the derivation in
Refs.~\cite{1,14}, it is
easily shown that if $p_i$ and $p'_i$ ($i=1$, 2, 3) are the incoming and
outgoing
momenta, then
\be
  p'_i = -p_i, \qquad i = 1,2,3.
\ee
On the other hand, if $d$ is odd then one has to introduce a symmetry operation
$T$ such that
\be
  T r = r, \qquad T \phi = {\pi\over 3d} - \phi,
\ee
so that $T$ transforms $0\leq \phi \leq \pi/3d$ into itself. One can easily show
that this $T$ operator, when applied to the angular eigenfunctions
$\Phi_l$ of the problem (see Eq.~(2.17c) of Ref.~[10]) yields
\be
  T \Phi_l = (-1)^l \Phi_l.
\ee
Following the steps in Refs.~\cite{10,14}, it is straightforward to prove that
$p'_i$ and $p_i$ are related by
\be \label{eq:p'-p}
  \pmatrix{p'_1\cr p'_2\cr p'_3\cr}
  = \pmatrix{0 &-a & b\cr -a & b & 0\cr b& 0 & -a\cr}
  \pmatrix{p_1\cr p_2\cr p_3\cr},
\ee
where
\be
  a = {\sin({\pi\over 3}-{\pi\over 3d})\over \sin (\pi/3)}, \qquad
  b = {\sin (\pi/3d)\over \sin(\pi/3)}.
\ee
It is interesting to note that the relation between $p'_i$ and $p_i$ is
similar to
that in the translation case~\cite{14}, except that $a$ and $b$ are
different in the
two cases. As expected, for $d = 1$,  one recovers the known relation
between $p'_i$
and $p_i$ for the Calogero model~\cite{1}.\par
%
%
A further generalization of Eqs.~(\ref{eq:genCMW-H}) and~(\ref{eq:gen-x-y}) is
also possible, and in that case $B^2$  is  given by $(0\leq\delta\leq \pi/6d)$
\be
  B^2 = - {\partial^2\over\partial\phi^2}
  + {9d^2g\over \sin^2(3d\phi+3d\delta)}.
\ee
One can now show that for even $d$, $p'_i = - p_i$, while for odd $d$,
$p'_i$ and
$p_i$ are related by Eq.~(\ref{eq:p'-p}), but where $a$ and $b$ are now given by
\be
  a = {\sin({\pi\over 3}-{\pi\over 3d}+2\delta)\over \sin (\pi/3)}, \qquad
  b = {\sin({\pi\over 3d}-2\delta)\over \sin (\pi/3)}.
\ee
For $\delta = \pi/6d$, we obtain $a=1$, $b=0$, so that $p'_1 = -p_2$, $p'_2 = -
p_1$,
$p'_3 = -p_3$ as in the CMW case~\cite{10}.\par
%
%
{}Finally, one can also consider a generalization of the CMW problem
characterized
by Eqs. (\ref{eq:newH}) and (\ref{eq:B}), but where
\be
  B^2 = - {\partial^2\over\partial\phi^2} + 9d^2\bigg [ {g\over
  \sin^2 3d\phi}+{f\over \cos^2 3d\phi}\bigg ],
\ee
with $d = 1$, 2, 3,~$\ldots$. In this case one can show that for every
integral $d$,
$p_i$ and $p'_i$ satisfy $p'_i=-p_i$ ($i=1$, 2, 3). A further
generalization consists
in replacing $\phi$ by $\phi +\delta$ ($0 \leq\delta\leq \pi/6d$), and one
can show
that irrespective of the values of $\delta$ and $d$, $p'_i = - p_i$.\par
%
%
Thus one has obtained a wide class of new exactly solvable three-body problems
where there is a simple but interesting relationship between $p'_i$ and
$p_i$. In all these cases, one can also add either the oscillator or the
Coulomb-like
potential~\cite{11}, and the full bound state problem is exactly solvable
in both the
cases.\par
%
%
One of us (AK) is grateful to Prof.~C.~Quesne for kind invitation and warm
hospitality during his stay at Universit\'e Libre de Bruxelles.\par
%
%

\end{document}